# Why Agentic-PRs Get Rejected: A Comparative Study of Coding Agents

Sota Nakashima
Kyushu University
Japan
nakashima@posl.ait.kyushu-u.ac.jp

Yuta Ishimoto
Kyushu University
Japan
ishimoto@posl.ait.kyushu-u.ac.jp

Masanari Kondo
Kyushu University
Japan
kondo@ait.kyushu-u.ac.jp

Shane McIntosh
University of Waterloo
Canada
shane.mcintosh@uwaterloo.ca

Yasutaka Kamei
Kyushu University
Japan
kamei@ait.kyushu-u.ac.jp

## Abstract

Agentic coding—software development workflows in which autonomous coding agents plan, implement, and submit code changes with minimal human involvement—is rapidly gaining traction. Prior work has shown that Pull Requests (PRs) produced using coding agents (Agentic-PRs) are accepted less often than PRs that are not labeled as agentic (Human-PRs). The rejection reasons for a single agent (Claude Code) have been explored, but a comparison of how rejection reasons differ between Agentic-PRs generated by different agents has not yet been performed. This comparison is important since different coding agents are often used for different purposes, which can lead to agent-specific failure patterns. In this paper, we inspect 654 rejected PRs from the AIDev dataset covering five coding agents, as well as a human baseline. Our results show that seven rejection modes occur only in Agentic-PRs, including distrust of AI-generated code. We also observe agent-specific patterns (e.g., automated withdrawal of inactive PRs by Devin), reflecting differences in how agents are configured and used in practice. Notably, a large proportion of rejected PRs (67.9%) lack explicit reviewer feedback, making their rejection reasons difficult to determine. To mitigate this issue, we propose a set of heuristics that reduce the proportion of such cases, offering a practical preprocessing step for future studies of PR rejection in agentic coding.

## 1 Introduction

*Agentic coding* [20] refers to software development workflows in which autonomous coding agents plan tasks, modify code, run tests, and iterate with minimal human involvement. Recent offerings, such as Claude Code, integrate well with social coding platforms like GitHub. These agents can submit Pull Requests (PRs),[1] which maintainers then review and either accept or reject.

Agentic-PRs are already being adopted at scale. Li et al. [15] released the AIDev dataset [14], a collection of 932,791 Agentic-PRs created between 2025-01-01 and 2025-08-01. Their analysis shows that the emergence of coding agents has substantially increased the volume of code contributions; however, Agentic-PRs are less likely to be accepted than PRs that are not labeled as agentic (*Human-PRs*). This suggests that current coding agents are not yet reliable enough to replace human developers.

In this paper, we compare the reasons why Agentic-PRs and Human-PRs are rejected to understand where state-of-the-art coding agents tend to underperform. By uncovering recurring failure patterns, our findings can support the early detection of common pitfalls and the development of self-improvement mechanisms for future agents.

Prior work [22] has explored the reasons for the rejection of PRs generated by Claude Code. We extend this line of work to the AIDev dataset, covering six types of PR authors that include Claude Code, OpenAI Codex, Devin, GitHub Copilot, and Cursor, as well as human developers. This broader setting allows us to compare rejection reasons between Agentic-PRs and Human-PRs, while also characterizing differences between agents. Since Li et al. [15] report that different coding agents tend to be used for different purposes, such multi-agent analysis is necessary to reveal agent-specific failure patterns that single-agent studies would miss.

More specifically, we set out to address the following Research Question (RQ):

*How do rejection reasons vary across coding agents?*

To address our RQ, we sample 654 rejected PRs from the dataset and inspect their review discussions to identify the rejection reasons. Our analysis, together with that of Watanabe et al. [22], reveals that a substantial portion of rejected PRs lack explicit reviewer feedback about why they were rejected. Their high prevalence would absorb the remaining rejection reasons and reduce the effective sample size for comparing the distribution of identifiable rejection reasons across coding agents, motivating us to propose a set of heuristics for filtering out such PRs.

In summary, this paper makes the following contributions:
- An empirically-grounded set of rejection modes that occur only in Agentic-PRs.
- A set of agent-specific rejection reasons.
- A set of heuristics for filtering out PRs whose rejection reasons cannot be determined from their review discussions, providing a practical preprocessing step for future studies.
- A detailed replication package providing the results of our inspection.[2]

---

[1] We refer to such pull requests as Agentic-PRs

[2] https://doi.org/10.5281/zenodo.18016038

## 2 Related Work

In this section, we discuss related work about agentic coding and PR rejection reasons.

### 2.1 Agentic Coding

The emergence of coding agents have pushed software development toward a new paradigm known as *agentic software engineering (SE 3.0)* [11, 12, 15]. In this context, agentic coding [20] refers to autonomous software development through coding agents capable of planning tasks, generating code, running tests, and committing changes to achieve a goal.[3] A growing body of empirical studies has examined how coding agents support or automate coding tasks [8, 15, 22]. Li et al. [15] published the AIDev dataset, a collection of 932,791 Agentic-PRs. Using this dataset, Horikawa et al. [13] conducted an empirical study on *agentic refactoring* and showed that 26.1% of commits produced by agents explicitly performed refactoring. Watanabe et al. [22] qualitatively analyzed 92 rejected PRs generated by Claude Code. They found that most rejections stemmed from project context, such as alternative solutions or PR size, rather than flaws in AI-generated code. Building on this line of work, our study examines rejection reasons for a more diverse set of PRs in the AIDev dataset, enabling us to compare rejection patterns between Agentic-PRs and Human-PRs, and to characterize differences across agents.

### 2.2 PR Rejection Reasons

In open-source software (OSS) development using version control systems such as GitHub, contributors typically propose changes to the codebase via PRs, and maintainers decide whether to merge or reject them. To understand development practices and support both contributors and maintainers, prior work [9, 16, 17, 21, 22] has investigated why PRs were rejected. Gousios et al. [9] analyzed 350 PRs that were closed without being merged and found that typical challenges of distributed collaboration, particularly coordination and communication issues, were more influential than code defects. Steinmacher et al. [21] complemented this perspective with a survey, highlighting duplication and misalignment with project goals as common reasons why teams closed PRs. Ogenrwor et al. [16, 17] identified key factors influencing AI-generated patch integration and PR rejection, including scope misalignment, maintainability concerns, redundant solutions, and procedural barriers such as incomplete documentation or administrative policies. Despite these studies, there has been no comparative investigation of how rejection reasons differ between Agentic-PRs and Human-PRs, nor of how these reasons vary across coding agents. As agentic coding becomes increasingly integrated into software development workflows, such a comparative view is essential for revealing the failure modes of autonomous agents and for designing mechanisms that improve their reliability and usefulness.

Table 1: Statistics of dataset

| PR type | #PR | Acceptance Rate |
|---|---|---|
| OpenAI Codex | 20,993 | 85.8% |
| Devin | 4,673 | 55.5% |
| GitHub Copilot | 3,891 | 55.0% |
| Cursor | 1,347 | 74.6% |
| Claude Code | 380 | 71.3% |
| Human | 6,149 | 82.6% |

## 3 Experimental Setup

In this section, we describe our dataset and study design.

### 3.1 AIDev Dataset

Our analysis is based on the AIDev dataset [14], which captures the contributions of autonomous coding agents to OSS development on GitHub. This dataset contains PRs created by five widely used agents (OpenAI Codex, Devin, GitHub Copilot, Cursor, and Claude Code), as well as PRs created by human developers. In this study, we focus on the AIDev-pop subset, which consists of repositories with at least 100 GitHub stars.[4] Following prior work [22], we exclude PRs that remain open (i.e., still under development), and consider merged PRs as *accepted* and closed-but-unmerged PRs as *rejected*. The statistics of the datasets are shown in Table 1.

### 3.2 Study Design

To answer our RQ, we adopt the classification framework proposed in prior work [19, 22], which identifies rejection patterns in refactoring PRs through the inspection of code review comments. We reuse the list of rejection categories defined by Watanabe et al. [22].
**Sample rejected PRs.** Since the number of rejected PRs varies across agents (Table 1), we stratify the data into six groups (five agents and Human-PRs). The smallest group is Claude Code, with 109 rejected PRs; therefore, we randomly sample 109 rejected PRs from each of the six groups, yielding 654 PRs in total.
**Code rejection reasons.** For each sampled PR, we read the associated review discussion and identify a primary reason for rejection. We first attempt to assign one of the rejection categories defined by Watanabe et al. [22]. When a rejection reason cannot be adequately captured by this list, we apply open coding to the rejection comments and inductively introduce a new category. The coding process was conducted by three annotators: each PR was independently reviewed by two annotators to ensure reliability, and disagreements were resolved through discussion.

## 4 Results

Table 2 shows the frequency of each rejection category based on the taxonomy proposed by Watanabe et al. [22]. Categories newly introduced in this study are shown in bold.

**Finding #1: Seven rejection reasons appear only in Agentic-PRs and highlight differences in how coding agents are used.** Table 2 shows that seven rejection categories are observed exclusively in Agentic-PRs, such as *No confidence in AI-generated code* (0.2% vs. 0.0%) and *Are too large* (1.5% vs. 0.0%). In one PR [1], the project owner closed the PR after criticizing the changes generated

---

[3]Agentic software engineering encompasses not only agentic coding but also broader collaborative activities involving multiple humans and agents [11].

[4]We use the latest version as of October 20, 2025.

Table 2: Rejection reasons for Agentic-PRs and Human-PRs.

| Category | Abbreviation | Description | %Agentic-PRs | %Human-PRs |
| --- | --- | --- | --- | --- |
| Are implemented by other PRs/developers | Alternative | The contributor or project team chooses a different solution before this PR can be merged. | 9.2% | 15.6% |
| Are inactive (author/community) | Inactive | A project's state of ceased or minimal development activity, often implying a lack of ongoing maintenance or community engagement. | 7.5% | 2.8% |
| **Submission for experimentation** | Experimentation | The PR is created to experiment with or evaluate a coding agent or development tool (e.g., testing its capabilities, prompts, or workflows) and is not intended for merging into the project. | 2.9% | 0.9% |
| **Wrong code location** | Wrong location | The PR changes are put in the wrong branch or repository of the project. | 1.8% | 1.8% |
| Contain choices of non-optimal design solutions | Non-optimal design | The PR implements a design that is considered suboptimal, inefficient, or architecturally unsound. | 0.9% | 5.5% |
| Are obsolete | Obsolete | The proposed changes become outdated or irrelevant due to evolving project requirements or newer implementations. | 0.7% | 5.5% |
| Are too large | Too large | The PR is too large or complex, making effective review impractical. | 1.5% | 0.0% |
| Not sure | Not sure | The PR has review comments, but they are ambiguous or open to multiple interpretations, making it difficult to classify the specific rejection reason. | 1.5% | 0.0% |
| Introduce bugs/ break APIs / breaks compatibility | Bug/API break | The PR introduces bugs into existing functionality, breaks API compatibility, or includes changes that disrupt normal system operation. | 1.1% | 0.9% |
| Are not in the community interest | Not in interest | The PR does not align with the project's direction or community goals, and is judged not worth the investment of limited resources. | 0.7% | 2.8% |
| Do not add value | No added value | The PR provides no clear or significant benefit to the project, its users, or its maintainers. | 1.1% | 0.0% |
| Introduce merge conflicts | Merge conflicts | Resolving the merge conflicts requires significant manual effort, beyond a simple rebase. | 0.7% | 0.9% |
| **Context/environment limitation** | Context limitation | The PR was not merged because AI coding agents could not access necessary information (e.g., private data and resources in separate repositories) required to complete the change. | 0.4% | 0.0% |
| **Deferred** | Deferred | The proposed change is delayed for further investigation in the future. | 0.4% | 0.0% |
| Submission for verification | Verification | The PR is created solely to trigger automated checks (e.g., CI pipelines) and is not intended for merging. | 0.2% | 0.9% |
| No confidence in AI-generated code | No AI confidence | The PR is rejected because the code was AI-generated and lacks sufficient human review or understanding to ensure reliability. | 0.2% | 0.0% |
| Increase complexity | More complex | The proposed solution was more complex than warranted. | 0.2% | 0.0% |
| Unknown (No feedback provided) | Unknown | The PR was closed without explanatory comments or discussion, preventing classification of the actual rejection reason. | 69.0% | 62.4% |

by Claude Code, stating "*I'll name and shame such contributions with the SLOP badge.*" A custom SLOP label was then applied to mark low-value AI-generated code. In another large PR [2], the contributor closed the PR after commenting "*Apologies for the large PR. I'll break this down into several smaller ones.*" It illustrates how oversized, multi-purpose Agentic-PRs can be rejected since they are difficult to review. Prior work [22] has shown that Agentic-PRs tend to be larger than Human-PRs in terms of the size and scope of the proposed changes. Consistent with this observation, in our sample all PRs rejected due to their size are Agentic-PRs, even though human developers can submit oversized PRs. These examples show that coding agents introduce rejection modes that do not appear for Human-PRs.

**Finding #2: Extending the dataset reveals agent-specific patterns and requires new rejection categories.** Figure 1 shows that PRs generated by Devin have a markedly higher proportion of rejections due to *Are inactive (author/community)* (32.1%). This behavior is consistent with the support in Devin for automatically closing inactive PRs [7]. For example, one PR [3] was closed after Devin commented "*Closing due to inactivity for more than 7 days.*" Additionally, we introduce four new categories (e.g., *Submission for experimentation*) in this analysis. For example, in one experimentation-only PR [4], the author wrote "*Just an experiment to see how copilot would handle this.*" The PR was then closed, indicating that it was never intended to be merged. These new codes allow us to describe failure modes that are specific to coding agents.

**Finding #3: A large proportion of rejected PRs lack explicit feedback, making their rejection reasons difficult to determine.** In our sample, the majority of rejected PRs (67.9%) do not include clear reviewer feedback about why the change was rejected; these are labeled as *Unknown (No feedback provided)*. Figure 1 shows that *Unknown* PRs are particularly frequent for PRs generated by OpenAI Codex, GitHub Copilot, and Cursor. These PRs provide no actionable information about the underlying rationale and act as noise for qualitative analysis aiming to identify rejection reasons. The high prevalence of *Unknown* PRs reduces the effective sample size and can obscure patterns in the remaining rejection categories. To mitigate this issue, we design and evaluate a simple heuristic filtering (Section 5) that aims to remove *Unknown* PRs.

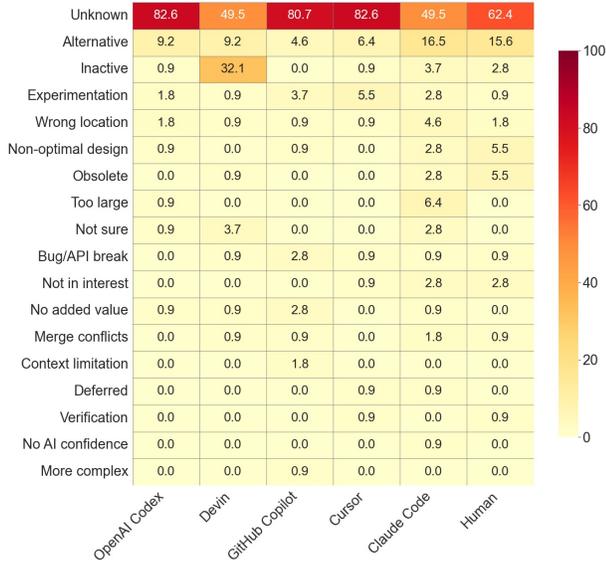

Figure 1: Distribution of rejection reasons by agent

> **Answer to RQ:** Rejection reasons differ between Human-PRs and Agentic-PRs as well as across coding agents.
> **Agent-only rejection modes.** Seven rejection categories (e.g., No AI confidence and Oversized PRs) appear only in Agentic-PRs.
> **Agent-specific patterns.** Our extended, multi-agent dataset reveals agent-specific profiles (e.g., automated withdrawal of inactive PRs by Devin) and motivates new categories (e.g., Experimentation-only PRs).
> **Limited feedback.** A large proportion of rejected PRs are labeled as *Unknown (No feedback provided)*, which constrains the effective sample size for analyzing rejection reasons.

## 5 Heuristic Filtering of *Unknown* PRs

**Motivation.** As shown in our manual analysis, *Unknown* PRs account for 67.9%. Since these PRs lack explicit feedback about why the change was rejected, they provide no reliable basis for assigning rejection codes, acting as noise in qualitative analysis. This observation motivates us to explore heuristics that can filter out such non-informative PRs and clean the dataset for future analysis. During the manual coding, we observed shared patterns in how such PRs were closed. For instance, the contributor opened and then self-closed the PR on the same day without any discussion [5]. In contrast, as an example of a *Non-Unknown* PR [6], the reviewer clearly explained why the change should not be merged and later closed the PR, providing a successful case in which the closer and feedback provider are the same person. We leverage these patterns to design our heuristics.

**Approach.** Our goal is to design initial heuristics that can filter out a substantial portion of non-informative PRs. Based on our manual coding experience, we employ three heuristics: (1) whether the PR is self-closed by its author, (2) whether the resolution time (from PR creation to closure) is short, and (3) whether the person who closes the PR has left no comment in the PR thread. To clarify the impact of these heuristics, we evaluate each heuristic and their combinations using our manually coded dataset of 654 rejected PRs.

Table 3: Performance of heuristics for filtering *Unknown* PRs

| Heuristic | Macro-Precision | Macro-Recall | Macro-F1 |
| --- | --- | --- | --- |
| Self-close only | 63.2% | 58.2% | 58.0% |
| Resolution Time only | 66.2% | 65.7% | 65.9% |
| No-comment only | 72.9% | 74.5% | 66.6% |
| **Resolution Time + Self-close** | 69.0% | 69.1% | **69.0%** |
| Self-close + No-comment | 71.3% | 71.1% | 61.8% |
| **Resolution Time + No-comment** | **73.0%** | 74.5% | 66.4% |
| All three | 71.7% | 71.3% | 61.9% |

Table 4: Recommended heuristics with their parameters

|  | Heuristics | Parameter setting |
| --- | --- | --- |
| Pattern A | Resolution Time<br>Self-close | ≤ 7 days<br>Yes |
| Pattern B | Resolution Time<br>No-comment | ≤ 45 days<br>Yes |

Following prior work [18, 23], we report macro-averaged precision, recall, and F1 to prevent the majority class (i.e., *Unknown* PRs) from dominating the score.

**Results.** We summarize the evaluation results in Table 3. Based on these results, we identify two recommended heuristics with their parameter settings (Table 4). When the goal is not only to reduce *Unknown* PRs but also to retain a sufficient number of *Non-Unknown* PRs, we recommend Pattern A, which achieves the highest macro-F1. In contrast, when the analysis requires conservative filtering that aims to avoid filtering out *Non-Unknown* PRs as much as possible, we recommend Pattern B, which achieves the highest macro-precision. Although these heuristics are simple, they effectively filter out *Unknown* PRs, providing a practical preprocessing step for future large-scale studies of PR rejection in agentic coding.

## 6 Threats to Validity

**Construct Validity.** Our analysis relies on manual coding of PR discussions. To mitigate subjectivity, each PR was independently labeled by two annotators, and any disagreements were resolved through discussion until a consensus was reached.

**Internal Validity.** Our quantitative comparisons of rejection codes (e.g., between Agentic-PRs and Human-PRs) rely on relative frequencies. Because the majority of rejected PRs are labeled as *Unknown*, the effective sample sizes for the remaining rejection categories are relatively small and may be insufficient to support strong claims about differences in their distributions.

**External Validity.** Our study focuses on decision making in pull requests, so the findings may not directly generalize to other kinds of project decisions. Hao et al. [10] has analyzed developers' shared conversations with ChatGPT in GitHub issues, suggesting that future work should extend our analysis to issue-based decision making involving autonomous coding agents.

## 7 Conclusion & Future Work

In this paper, we qualitatively investigate why Agentic-PRs and Human-PRs are rejected in the AIDev dataset. By inspecting 654 rejected PRs across five coding agents and a human baseline, we uncover rejection modes that occur only in Agentic-PRs (e.g., distrust of AI-generated code) and agent-specific patterns (e.g., automated withdrawal of inactive PRs by Devin). We also find that most rejected PRs (67.9%) lack clear reviewer feedback, which reduces the effective sample size for analyzing rejection reasons and motivated our design of simple heuristics to filter out such cases.

Based on these findings, we suggest the following two research directions. First, our heuristic filtering of PRs without identifiable rejection reasons should be applied to larger samples from AIDev and similar datasets, enabling more robust comparisons of rejection reason distributions with sufficient sample sizes. Second, while this study focuses on explicit rejection reasons recorded in review discussions, future work should investigate implicit signals that may shape rejection decisions (e.g., test failures and code churn) and relate these quantitative factors to acceptance outcomes to obtain a more comprehensive view of how agentic contributions are evaluated in practice.

## Acknowledgement

We gratefully acknowledge the financial support of: (1) JSPS for the KAKENHI grants (JP24K02921, JP25K03100, JP25K22845); (2) Japan Science and Technology Agency (JST) as part of Adopting Sustainable Partnerships for Innovative Research Ecosystem (ASPIRE), Grant Number JPMJAP2415, and (3) the Inamori Research Institute for Science for supporting Yasutaka Kamei via the InaRIS Fellowship.